# 双编码特征过滤泛化注意力 UNET 视网膜血管分割

陶西洛[1], 吴达文[2], 唐青青[2], 赵凯洋[3], 尹 腾[1],
李彦霏[1], 尚文一[1], 刘晶玉[1], 张海仙[1]
(1. 四川大学计算机学院机器智能实验室, 成都 610064; 2. 四川大学华西医院眼科, 成都 610041;
3. 四川大学华西临床医学院, 成都 610041)

**摘 要:** 视网膜血管的形态与功能改变对临床中眼部及心血管疾病的诊断具有重要指导意义. 依靠人工观察和经验判断眼底图像中视网膜血管状态不仅效率低下且主观性强,因此智能视网膜血管分割技术应运而生. 随着深度学习技术不断发展,深度神经网络模型显著提高了视网膜血管的分割性能,但该领域依旧面临着有限的训练数据、数据分布不平衡和特征提取不充分等问题,阻碍了分割性能和模型泛化能力的提升. 为了解决这些关键问题,本文提出了 DEFFA-Unet,其通过增加专用编码器来处理域不变的预处理输入,以增强特征编码的深度和模型的泛化能力. 开发了一个特征过滤融合模块,以确保精确的特征过滤和稳定的特征融合. 针对高精度、低误诊率要求的临床应用场景,传统的跳跃连接被替换为注意力引导的特征重构融合模块. 此外,本文提出了一种基于目标特征相似性的独特平衡方法 JESB,以应对数据分布不平衡的问题;并提出了一种创新的数据增强技术 SOTA-CSA,以解决数据限制和跨域问题. 实验在 4 个基准数据集(DRIVE、CHASEDB1、STARE 和 HRF)和一个 SLO 数据集(IOSTAR)上进行,采用全面的评估指标进行验证. 实验结果表明了 DEFFA-Unet 相较于基线模型和最先进模型的优越性.

# Dual encoding feature filtering generalized attention UNET for retinal vessel segmentation

*ISLAM Md Tauhidul*[1], *WU Da-Wen*[2], *TANG Qing-Qing*[2], *ZHAO Kai-Yang*[3],
*YIN Teng*[1], *LI Yan-Fei*[1], *SHANG Wen-Yi*[1], *LIU Jing-Yu*[1], *ZHANG Hai-Xian*[1]
(1. Machine Intelligence Laboratory, College of Computer Science, Sichuan University, Chengdu 610064, China;
2. Department of Ophthalmology, West China Hospital, Sichuan University, Chengdu 610041, China;
3. West China School of Medicine, Sichuan University, Chengdu 610041, China)

**Abstract:** Retinal blood vessel segmentation is crucial for diagnosing ocular and cardiovascular diseases. Although the introduction of U-Net in 2015 by Olaf Ronneberger significantly advanced this field, yet issues like limited training data, imbalance data distribution, and inadequate feature extraction persist, hindering both

**收稿日期:** 2024-03-12 **修回日期:** 2024-04-11 **接受日期:** 2024-05-07
**基金项目:** 四川省科技厅重点研发项目(2022ZDZX0023)
**作者简介:** 陶西洛(1997—), 男, 硕士研究生, 研究方向为医学图像处理. E-mail: tauhidscu16@gmail.com
**通讯作者:** 张海仙. E-mail: zhanghaixian@scu.edu.cn

the segmentation performance and optimal model generalization. Addressing these critical issues, the DEFFA-Unet is proposed featuring an additional encoder to process domain-invariant pre-processed inputs, thereby improving both richer feature encoding and enhanced model generalization. A feature filtering fusion module is developed to ensure the precise feature filtering and robust hybrid feature fusion. In response to the task-specific need for higher precision where false positives are very costly, traditional skip connections are replaced with the attention-guided feature reconstructing fusion module. Additionally, innovative data augmentation and balancing methods are proposed to counter data scarcity and distribution imbalance, further boosting the robustness and generalization of the model. With a comprehensive suite of evaluation metrics, extensive validations on four benchmark datasets (DRIVE, CHASEDB1, STARE, and HRF) and an SLO dataset (IOSTAR), demonstrate the proposed method's superiority over both baseline and state-of-the-art models. Particularly the proposed method significantly outperforms the compared methods in cross-validation model generalization.

## 1 Introduction

The increasing incidence of diabetes and cardiovascular diseases substantially raises the likelihood of vision impairment, adding to the worldwide impact of ocular disorders like Diabetic Retinopathy (DR), glaucoma, and macular degeneration[1]. This situation underlines the vision impairment faced by roughly 2.2 billion people worldwide, as reported by the World Health Organization[2], with nearly half of these cases potentially preventable or treatable through enhanced diagnostic accessibility. Hence, automated Computer-Aided Diagnosis (CAD) systems play a vital role in early detection of diseases and addressing the worldwide challenge of vision loss. Utilizing fundus images, CAD system significantly advances disease detection, standardizes image analysis, and reduces observer variability[3], thereby meeting a vital need in healthcare.

The emergence of Deep Learning (DL) techniques, particularly Convolutional Neural Networks (CNNs), has markedly propelled progress in image classification and segmentation. Their inherent ability to extract and learn hierarchical feature representations from data makes them particularly effective for these tasks, offering a robust approach to automatically identifying relevant features for accurate classification and segmentation. Transitioning from layer-centric CNN approaches (such as the 5-Layer CNN by Fan *et al.* and the 6-Layer CNN by Liskowski *et al.*), which incorporate pre and post-processing techniques (including global contrast normalization, zero-phase component analysis, and adaptive histogram equalization), to more complex hybrid CNN models [4-9], a range of advanced CNN frameworks have emerged. These models effectively blend random forests, Conditional Random Fields (CRF), and fully convolutional networks, marking significant advancements in the field. Notably with the introduction of UNet[10], it has been widely adopted divide and conquer based technique and currently considered as baseline models for medical image segmentation tasks. With that, Jin *et al.* proposed the Deformable U-Net (DUNet)[11], merging the U-Net framework with deformable convolution blocks to enhance the model's ability to adapt to intricate vessel structures. Additionally, Guo *et al.* unveiled DRNET[12], employing a residual deep network design for advanced high-level feature reconstruction. Zhou *et al.* introduced more advanced architecture as UNet++[13] with subsequent dense convolutional block in skip connection to address the semantic gap between low-level and high-level feature map in feature reconstruction and later Wang *et al.* [14]combined patch learning with Nested-UNet. Veesle-Net[15] integrated the inception resnet[16] block in UNet structure for multi scale feature extraction potentially improved complex ves-sel segmentation. To address the smooth feature reconstruction Mou *et al.* [17] proposed attention based UNet, where features are filtered and highlighted for robust feature extraction through channel and spatial attention mechanism. Wu *et al.* [18] introduced NFN+, incorporating a cascaded approach with a secondary U-Net designed to refine initially misclassified vessels via a recursive learning methodology. Building on this foundation, Guo[19] furthered the cascaded model concept through the development of CSGNet, which leverages a feature-guided training strategy to specifically enhance the segmentation of thin vessels in areas of low resolution. With an extension, cascaded W-Net[20] additionally addressed the domain shifting and introduced potential technique for model generalization on unseen data. Most recently, Tan *et al.* [21] has advanced the cascaded U-Net paradigm by proposing a multi-task model to focus on thin vessels and vessels in lesion areas. To enhance the segmentation performance further and ensure the high-level feature extraction, dual encoding variance of UNet, CSU-Net[22] is introduced by Wang *et al*. In the context recently, Li *et al*, guided the dual encoder for local and global feature extractions in Ref. [23]. Subsequently, multi-decoder based HANet[24] is also introduced to focus vessel in complex conditions. In the similar manner, variance like multi-modal with representative input features, and multi-task with task specific decoders are also explored in Refs. [25] and [26] for comprehensive vessel segmentation.

However, still there are several challenges to address. Firstly, these methods often rely on one-way conventional feature encoding, which restrict the richer feature extractions. Secondly, many of these methods are designed with a specific domain in mind, limiting their performance and adaptability across different data domains. Consequently, the issue of generalization remains a significant hurdle yet to be overcome [27,28]. In addressing above issues, we propose the DEFFA-UNet. This model is distinctively structured based on dual encoding mechanisms to capture both domain-specific ($D_{\text{specific}}$) and domain-invariant ($D_{\text{invariant}}$) high-frequency guided vessel features from raw and preprocessd fundus image. Accordingly, a comprehensive Feature Filtering Fusion (FFF) module is designed to ensure robust hybrid feature fusion. Furthermore, traditional skip connection is replaced with Feature Reconstructing Fusion (FRF) module to address both the lower precision issue and generalized-feature reconstruction. In addition, this framework sets the stage for addressing data imbalance and data limitation with novel methods. These steps collectively bolster both the model' s segmentation performance and ability to generalize across varied and unseen data, marking a significant leap forward in the field of fundus image segmentation. Our contributions could be summarized as:

(1) A dual encoding network is proposed with hybrid feature encoding novel method to ensure both richer feature extraction and enhanced generalization capabilities.

(2) A Feature Filtering Fusion (FFF) module is proposed to incorporate both channel and spatial attention mechanisms at the bottleneck preceding layer, ensuring precise feature fusion for subsequent layers.

(3) A Feature Reconstructing Fusion (FRF) module is introduced by utilizing novel attention method and multi-scale feature fusion mechanism to ensure high-precision and generalized feature- reconstruction.

(4) A unique data balancing method called jaccard-enhanced synthetic balancing (JESB) based on target feature similarities and an innovative data augmentation technique called SOTA-CSA are proposed to address data limitations and domain variance.

(5) The proposed approach is rigorously tested across five datasets, showcasing its superior performance in vessel segmentation and generalization capabilities.

## 2 Materials and methods

This section describes the detailed workflow and methodology employed in our study. As depicted in Fig. 1, our approach is bifurcated into two primary segments: data processing and the model designing phase. The data processing phase encompasses precise data balancing, data-augmentation and processing of domain-invariant data to ensure robust model training. The modeling phase is characterized by the implementation of a dual-encoder UNet architecture, which incorporates a Feature Filtering Fusion module, a modified res-inception decoder, and a feature reconstructing fusion module. The model architecture is shown in Fig. 3. Each of modular designs is elaborated upon in the subsequent sections, providing a detailed insight into the methodologies adopted to achieve our research objectives.

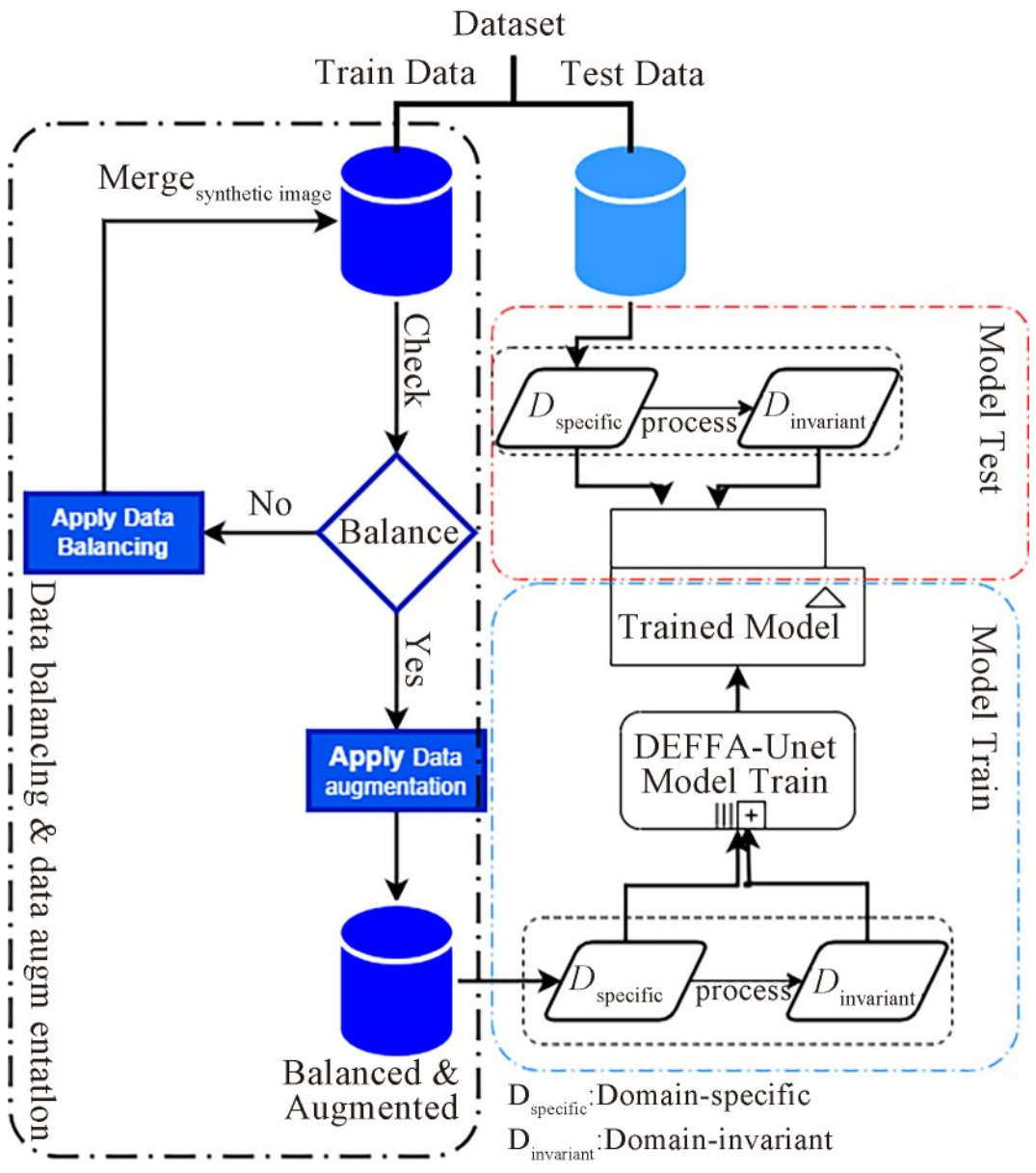

Fig. 1　Workflow of the proposed DEFFA-Unet

## 2.1 Dataset overview

In recognition of the pronounced challenges associated with obtaining annotated-data within this domain, an extensive literature review is undertaken to ensure high-quality data acquisition and found that DRIVE[29], STARE[30], CHASEDB1[31], and HRF[32] datasets are prevalently employed as benchmarks for evaluating retinal vessel segmentation algorithms. Notably, the DRIVE dataset is considered to be more standard for its distinct training and testing split facilitating uniform and comparable results. Therefore, we integrates these four foundational datasets with the newly available IOSTAR[33] dataset to have an extensive assessment against traditional and modern standards. Tab. 1 summarizes the experiment datasets where ‘FOV’ denotes Field Of View, ‘H’ denotes Healthy-sample and ‘R’ denotes retinopathy-diagnosed samples.

In order to align with our experimental framework and available computational resources, training data is re-sized and detailed in Tab. 1. It should be noted that the FOV masks were unavailable for CHASE[31] and STARE[30] datasets. Therefore, we adopted the FOV masks generated by the W-Net[20] and only the pixels within FOV masks is considered in the evaluation process.

Tab. 1　Experimental datasets overview

| Db name | Resolution | FOV | Total data | FOV mask | Resized |
|---|---|---|---|---|---|
| DRIVE[29] | 565 × 584 | 45° | 40, R:7, H:33 | Available | 568 × 584 |
| STARE[30] | 605 × 700 | 35° | 20, R:10, H:10 | Not availabl | 608 × 704 |
| CHASE[31] | 999 × 960 | 30° | 28, R:28, H:0 | Not available | 1000 × 960 |
| HRF[32] | 3504 × 2336 | 45° | 45, R:30, H:15 | Available | 512 × 512 |
| IOSTAR[33] | 1024 × 1024 | 45° | 30, R:0, H:30 | Available | 768 × 768 |

### 2.1.1 Data balancing with Jaccard-Enhanced Synthetic Balancing

As shown in Tab. 1, experiment datasets- DRIVE, CHSE, HRF and IOSTAR are highly imbalanced in healthy and retinopathy sample distribution which poses a risk of bias model-training. Inspired by the KMeans SMOTE technique proposed by H Hairani *et al.* in Ref. [34], we introduce Jaccard- Enhanced Synthetic Balancing (JESB) utilizing Jaccard Distance to assess similarities in binary label masks and implements KMeans clustering. To the best of our knowledge, we are the first to utilize this method in retinal fundus image segmentation task. This clus-tering optimized by silhouette scores, determines the optimal number of clusters ($K$) using Jaccard Distance metrics across samples. Specifically, for a collection of bi-nary images, $B_{i\,i=1}^{N}$, where $N$ is the total image count and $B_i$ denotes the binary mask of the $i^{th}$ image, the Jaccard distance between any two images $B_i$ and $B_j$ is calculated in Equation (1).

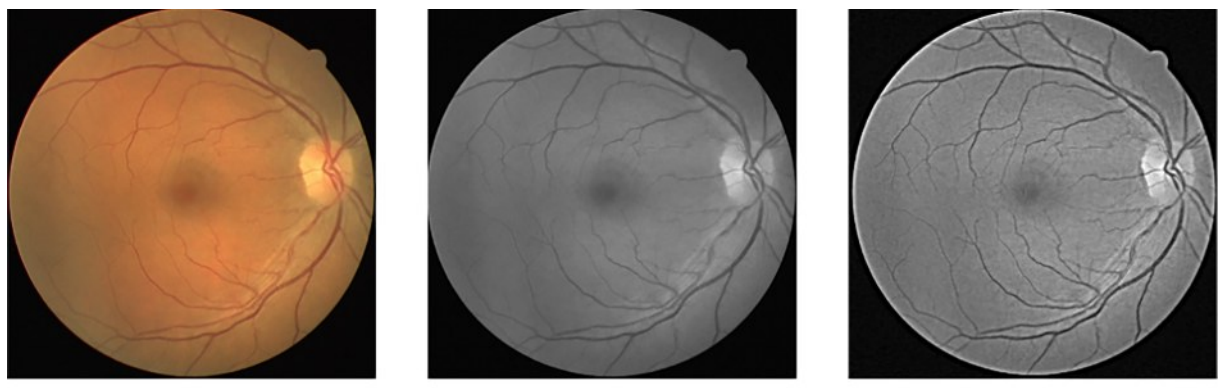

Fig. 2　Domain-invariant image processing

$$J\left(B_i,B_j\right)=1-\frac{\left|B_i\cap B_j\right|}{\left|B_i\cup B_j\right|} \tag{1}$$

The ideal cluster count, $k^*$ is identified by optimizing the silhouette score, s across a range of $k$ values. Here, $s$ is defined as $s=\frac{b-a}{\max\left(a,b\right)}$, where $a$ refers to the average distance within clusters, and $b$ represents the average distance to the nearest cluster. This is expressed as:

$$k^*=\underset{k}{\operatorname{argmax}}\, s\left(k\right),\ k=2,3,\cdots,K_{\max} \tag{2}$$

Finally, let $C=C_1,C_2,\cdots,C_{k^*}$ represent the clusters obtained. The target size $T$ for balancing is $T=\max\left(\left|C_1\right|,\left|C_2\right|,\cdots,\left|C_{k^*}\right|\right)$. For clusters $C_i$ with $C_i<T$, generate $G_i=T-\left|C_i\right|$ synthetic images with Color shifting, sharpening, and contrast changing augmentations following the method in Ref. [27]. The balanced dataset $D$ combines original and synthetic images $(S_{i1},S_{i2},\cdots,S_{iG_i})$ across all $k^*$ clusters:

$$D=\bigcup_{i=1}^{k^*}\left(C_i\cup S_{i1},S_{i2},\cdots,S_{iG_i}\right) \tag{3}$$

2.1.2　Data augmentation with SOTA-CSA　In medical imaging, particularly for retinal vessel segmentation, the necessity for data augmentation stems from the limited data availability. Traditional GAN-based methods, despite enhancing model performance, necessitate substantial computational resources. Drawing inspiration from the work of Tan *et al.* [21] and Qi *et al.* [35], our approach, termed SOTA-CSA, leverages color space mixing and normalization-guided augmentation strategies. This method capitalizes on the statistical properties of color spaces- namely the mean ($\mu_c$) and standard deviation ($\delta_C$) within State-Of-The-Art(SOTA) datasets. By utilizing SOTA datasets as reference, proposed method ensure both data consistency and data diversity which enhance model generalization on unseen data.

Mathematically, the calculation of $\mu_c$ and $\delta_C$ for color channels, $C\in R,G,B$ within SOTA dataset, $D$ is calculated following the Equations (4) and (5) where the total pixel count is $N$, the image count is $n$, and source pixel's value is $p_{ijk}^{C}$. The augmentation is applied based on calculated SOTA color statistics on the source dataset, $S$ in Equation(6). Augmented images are blended with original images using a randomly generated blending factor, $\alpha$ ($\alpha\in[0.7,1.0]$) and the augmentation is further enriched by applying random rotation, thereby introducing additional variability into the training set.

$$\mu_C=\frac{1}{N}\sum_{ijk}p_{ijk}^{C} \tag{4}$$

$$\sigma_C=\sqrt{\frac{1}{N}\sum_{ijk}\left(p_{ijk}^{C}-\mu_C\right)^2} \tag{5}$$

$$p_{ijk,C}^{\mathrm{aug}}=\alpha\left(\frac{p_{ijk}^{C}-\mu_C}{\sigma_C}\right)+\left(1-\alpha\right)p_{ijk}^{C} \tag{6}$$

2.1.3　Domain-invariant data process　Existing domain generalization methods [36,37] mainly focused on adversarial domain augmentation which lacks end-to-end train-ability therefore more focused has been paid towards domain adaptation techniques that leverage frequency-based features, particularly Fourier Domain Adaptation (FDA)[38]. This methods employ fast Fourier transformations to blend high and low-frequency elements across domains for enhanced generalization where mainly domain variant features are interchanged and randomized [39]. Although this methods improve model generalization, it still needs the multi-domain data in feature randomization which highlight the ongoing struggle to achieve optimal solutions in tasks like retinal vessel segmentation due to the of target domain and additional source domain data. Drawing inspiration from FDA based method, we utilized the power of UNet architecture, featuring a dual encoding method to extract features from both domain-specific and domain-invariant inputs where domain-invariant input are derived following FDA techniques. Our strategy centers solely on the source domain, deliberately avoiding the incorporation of target domain features in the randomization process. Instead, it highlights domain-invariant vessel features, steering the network towards these consistent elements. This guidance helps the network preserve essential vessel features when encountering unseen target data

Our method focuses on enhancing high-frequency components ($H$) from retinal images ($I$) through a dual-phase approach. Initially, we compute the local average ($L_{avg}$) to isolate low-frequency elements, which are then subtracted from $I$ to extract $H$. This procedure, aimed at amplifying vascular feature-relevant high-frequency details, is succinctly expressed mathematically as follows:

$$L_{avg}(x,y)=\frac{1}{\text{window_size}^2}\sum_{i=-n}^{n}\sum_{j=-n}^{n}+ I(x+i,y+j) \tag{7}$$

$$H(x,y)=I(x,y)-L_{avg}(x,y) \tag{8}$$

Where window_size denotes the dimension of the local averaging window, and $n$ is half of window_size, ensuring a localized averaging effect.

Following the extraction of high-frequency components, an enhancement factor ($G$) is computed to facilitate the selective amplification of these components. The enhancement factor is derived from the global variance ($\delta_x^2$) and the mean square error of $H$, as shown below:

$$\sigma_x^2=\frac{1}{M\times N}\sum_{x,y}(H(x,y)-\bar{H})^2$$
$$G=\alpha\times\frac{\sum_{x,y}H(x,y)^2}{M\times N(\sigma_x^2+\epsilon)} \tag{9}$$

In the Equation (9), $\alpha$ is utilized as a scaling factor to modulate the enhancement degree, $M\times N$ signifies the image dimensions, is the average of $H$, and $\epsilon$ serves as a minimal constant to prevent division by zero.

The high-frequency components are finally enhanced using the calculated factor $G$, rendering the domain-invariant vessel features more prominent:

$$I_{enhanced}(x,y)=G\times H(x,y) \tag{10}$$

The process is visualized in Fig. 2 where Fig. 2a presents an original image, Fig. 2b showcases the green channel image and Fig. 2c showcases the domain-invariant pre-processed image.

### 2.2 Dual encoder unet

Drawing inspiration from the base UNet[10] architecture, an encoder-decoder based architecture is chosen. Inspired by the dual encoding Unet architecture in Refs. [22, 23] and considering lightweight version in Refs. [20], a lightweight dual encoding baseline model is designed.

At first three building blocks namely-Base Block (BB), sequential-incremental block (SIB) and Stacked Generalization Block(SGB) is designed as shown in Equations (11) (12) and (13). Here the BB is designed following the basic building block of baseline UNet with DropBlock[40] regularization, the SIB block is designed to grasp intricate features by expanding the receptive field and the SGB block aims to achieve a denser and more effective representation of input data through channel-wise feature integration, preserving the spatial dimensions intact.

$$Y_{BB}=\text{ReLU}(\text{BN}(\text{DropBlock}(\text{Conv}_{3\times3}(X)))) \tag{11}$$

$$Y_{SIB}=\text{ReLU}\left(\text{BN}\left(\begin{array}{c}\text{DropBlock}\\(\text{Conv}_{3\times3}(\text{Conv}_{3\times3}(X)))\end{array}\right)\right) \tag{12}$$

$$Y_{SGB}=\text{ReLU}\left(\text{BN}\left(\begin{array}{c}\text{DropBlock}\\(\text{Conv}_{3\times3}(\text{Conv}_{1\times1}(X)))\end{array}\right)\right) \tag{13}$$

Then the original UNet encoder is replaced with two distinct encoder and a decoder where Encoder-1 integrates two pivotal blocks as shown in Equation (14) to extract domain-specific, $D_{specific}$ complex features. Encoder-2, is designed to extract domain-invariant, $D_{invariant}$ features with a sequence of BB and SGB block as shown in Equation (15). Accordingly, a connect block with two base block in Equation (16) and a modified res-inception decoder is designed with a parallel convolution block and residual connections as shown in Equation (17):

$$D_{\text{specific}} = Y_{\text{SIB}}\left(Y_{\text{BB}}(X)\right) \tag{14}$$

$$D_{\text{invariant}} = Y_{\text{SGB}}\left(Y_{\text{BB}}(X)\right) \tag{15}$$

$$C_{\text{b}} = Y_{\text{BB}}\left(Y_{\text{BB}}(X)\right) \tag{16}$$

$$R_I(X) = ReLU\left(\text{BN}\left(\text{Conv}_{1\times 1}(X)\right)\right) + Y_{\text{SGB}}\left(Y_{\text{BB}}(X)\right) \tag{17}$$

Finally, The original UNet architecture is enhanced by incorporating a three-layered dual encoder with channel dimensions [16, 32, 64] and decoder blocks, interconnected through a connect block, to construct the dual-encoder Unet.

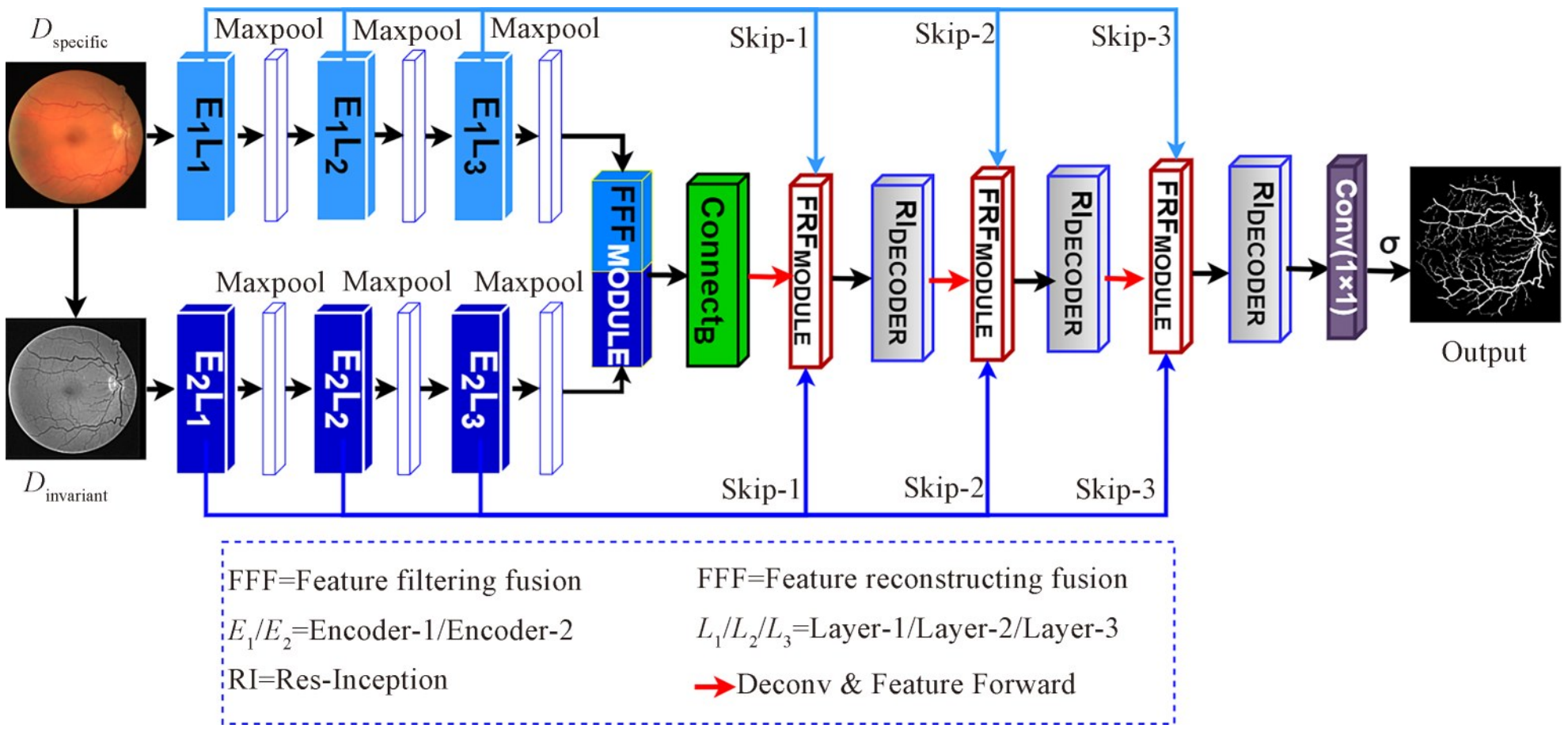

Fig. 3　DEFFA-Unet framework

## 2.3 Feature filtering fusion module

Backbone dual encoder model with $D_{\text{specific}}$ and $D_{\text{invariant}}$ inputs ensures richer feature extraction. Inspired by the contextual attention mechanisms of CBAM [41] and the channel-spatial interplay in CS Net [16], A Feature Filtering Fusion (FFF) module is designed as shown in Fig. 4. This module is adept at discerning 'what' features to emphasize within the feature maps and 'where' to allocate focus.

The FFF module functions by processing intermediate feature maps derived from both encoders. Let $F_{x_3}$ symbolize the domain-specific intermediate feature map, while $F_{y_3}$ signifies the domain-invariant feature map, with both $F_{x_3}$ and $F_{y_3}$ belonging to $R^{C\times H\times W}$, where $C$, $H$, and $W$ denote the dimensions of channel, height, and width, respectively.

'What' to Focus: To extract essential features from $F_{x_3}$, channel attention (CA) mechanisms are utilized, leveraging both the global average-pooled, ($F_{x_3}^{C_{\text{avg}}}$) and max-pooled, ($F_{x_3}^{C_{\max}}$) feature maps to capture comprehensive spatial details across channels. Through the application of a shared Multi-Layer Perceptron (MLP), we derive two separate channel-wise attention maps. These maps are then combined through element-wise summation to produce $F_{x_3}^{C_m}$. A sigmoid function applied to $F_{x_3}^{C_m}$ yields the attention weights, which are applied to domain-specific intermediate feature map via element-wise multiplication, resulting in the refined feature map $M_C(F_{x_3})$. This process is encapsulated in Equation (18).

$$M_C(F_{x_3}) = F_{x_3} \odot \sigma\left(\text{MLP}\left(\frac{1}{H\times W}\sum_{h=1}^{H}\sum_{w=1}^{W} F_{x_3}(c,h,w)\right) + \text{MLP}\left(\max_{h=1}\max_{w=1}^{W} F_{x_3}(c,h,w)\right)\right) \tag{18}$$

'Where' to Focus: Aligning with our objective to accentuate high-frequency domain-invariant features, Spatial Attention (SA) is applied to $F_{y_3}$. This process entails the retention of features ob-tained through average and max pooled feature maps ($F_{y_3}^{S_{avg}}$, $F_{y_3}^{S_{max}}$) followed by their concatenation into $F_{y_3}^{s_c}$. Sub-sequently, this concatenated feature map is refined using a (7×7) convolution, resulting in the extraction of comprehensive spatial features. Then, the spatial attention weights are derived with sigmoid function and applied on the intermediate feature, $F_{y_3}$ to enhance the prominence of distinct spatial features as delineated in Equation (19). This method effectively suppresses the low-frequency noises, ensuring the focus is maintained on the most relevant spatial vessel characteristics.

$$
\begin{aligned}
&F_{y3}^{s_{avg}} = \frac{1}{C}\sum_{c=1}^{C} F_{y_3}(c,h,w) \\
&F_{y3}^{s_{max}} = \max_{c=1}^{C} F_{y_3}(c,h,w) \\
&F_{y3}^{sc1} = \mathrm{Conv}_{7\times 7}\left(\mathrm{concat}\left(F_{y_3}^{s_{avg}}, F_{y_3}^{s_{max}}\right)\right) \\
&M_s\left(F_{y_3}\right) = F_{y_3} \odot \sigma\left(F_{y_3}^{s_{c1}}\right)
\end{aligned} \tag{19}
$$

The CA and SA filtered features are then fused to construct a comprehensive hybrid feature set. This fusion process is further refined through convolution, batch-normalization, and ReLU activation as shown in Equation(20) to ensure a cohesive integration of feature .

$$
F_{fused} = \mathrm{ReLU}(\mathrm{BN}(\mathrm{Conv}(\mathrm{concat}\left(M_C(F_{x_3}), M_s(F_{y_3})\right)))) \tag{20}
$$

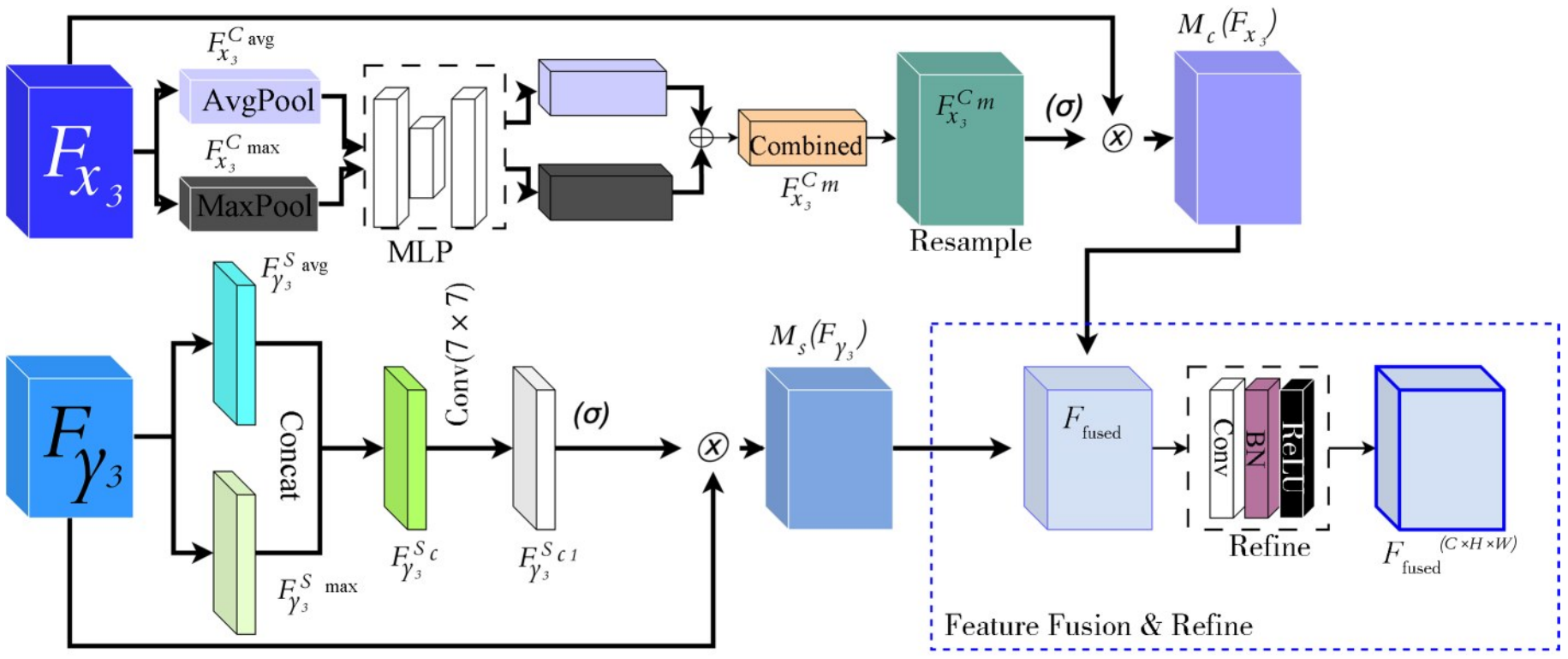

Fig. 4　Feature filtering fusion module

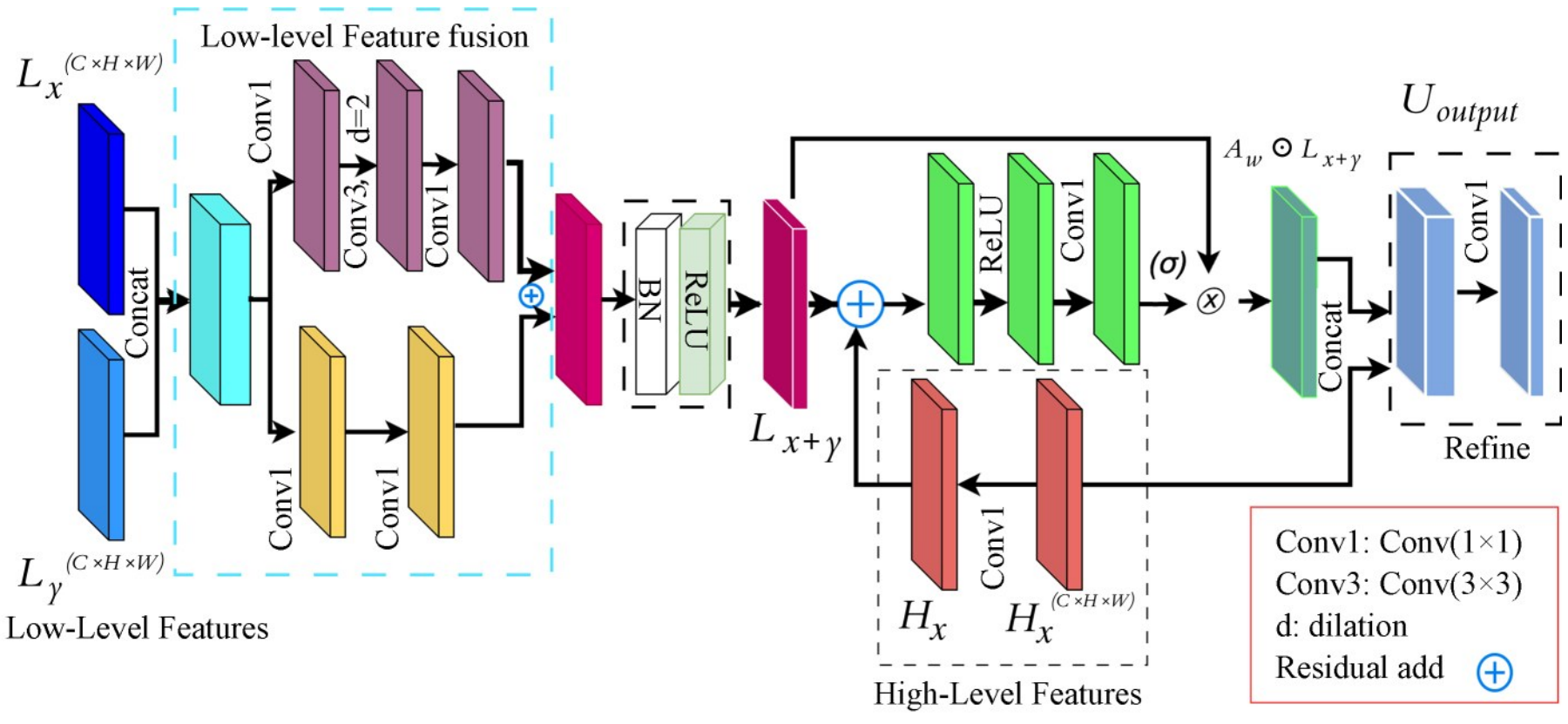

Fig. 5　Feature reconstructing fusion module

## 2.4　Feature reconstructing fusion module

Traditional U-Net's skip connections, while facilitating feature fusion, can inadvertently cause semantic mismatches due to direct feature reconstructing methods. Addressing the issue Oktay *et al.* [42] introduced multi-level attention gate for low-level feature processing. Motivated by this, feature reconstructing fusion(FRF) module is designed. Instead of leveraging only low-level features, this module utilized both low-level and high-level features to generate attention weights as illustrated in Fig. 5. In the figure, $L_x$ and $L_y$ are low-level, D subscript (specifc) and D subscript (invariant) feature maps, each characterized by their spatial dimensions $C \times H \times W$. The FRF Module first operates on low-level features-fusing by concatenating $D_{specific}$ and $D_{invariant}$ feature maps which undergoes through two parallel path: one with dilated convolution and projection layer to capture the broader context by filtering irrelevant features and the other with projection and convolution layer to capture relevant localized features. Then both feature sets are combined and refined as shown in Equation (21) ensuring high precision with generalize-able feature map $L_{x+y}$.

$$L_{x+y} = \mathrm{ReLU}\Big(\mathrm{BN}(\mathrm{Conv}_{1\times1}\Big) \mathrm{Conv}_{3\times3}^{dil=2}\Big(\mathrm{Conv}_{1\times1}\Big(\mathrm{concat}\big(L_x, L_y\big)\Big)\Big) + \mathrm{Conv}_{3\times3}\Big(\mathrm{Conv}_{1\times1}\Big(\mathrm{concat}\big(L_x, L_y\big)\Big)\Big) \tag{21}$$

Furthermore, as shown in Fig. 5 , both the low level and high-level features are fused at the adjacent scale to generate comprehensive attention weights, $A_w$. The attention is applied on fused low-level feature $L_{x+y}$ to scale each feature's intensity based on feature importance. Finally, the features augmented with attention are merged with the high-level feature map $H_x$ to allow the module to integrate detailed feature information with broader contextual cues. The channel dimensions are adjusted for subsequent processing with a final 1×1 convolution. The detailed procedure of attended feature reconstruction can be expressed as:

$$A_w = \mathrm{Sigmoid}\Big(\mathrm{Conv}_{1\times1}\Big(\mathrm{ReLU}(\mathrm{Conv}_{1\times1}(H_x) + L_{x+y}))\Big) \tag{22}$$

$$U_{output} = \mathrm{Conv}_{1\times1}\Big(\mathrm{concat}\big(H_x, A_w \odot L_{x+y}\big)\Big).$$

# 3 Experiments

This section will introduce the experiment set up, evaluation metrics and result evaluation. Detailed description will be given in subsections.

## 3.1 Environment setup

In our research, outlined in Tab. 1, we employed five publicly available datasets. Notably, only the DRIVE dataset came with predefined training and testing partitions. In alignment with methodologies established in previous studies [20, 21], we distributed the other datasets into specific training and testing sets as follows: CHASEDB1 was divided into 20 training images and 8 testing images, STARE was allocated 10 images for training and another 10 for testing, and IOSTAR was organized into 25 training images and 5 testing images. Additionally, for cross-domain testing, we utilized the entire collection of images from all datasets to ensure a comprehensive assessment.

Given the challenges posed by class imbalance and the varying thicknesses of vessel structures, a weighted combination of Binary Cross-Entropy (BCE) [43] and Dice loss [44] functions is utilized. The Binary Cross-Entropy (BCE) Loss is determined in Equation (23) where $pr_i$ signifies the predicted probability for the $i^{th}$ pixel, $gt_i$ is the actual label for the $i^{th}$ pixel, and $N$ denotes the image's total pixel count.

$$\mathrm{BCE}\big(pr, gt\big) = -\frac{1}{N}\sum_{i=1}^{N}\Big[gt_i \cdot \log\big(pr_i\big)\Big) + (1 - gt_i) \cdot \log(1 - pr_i)\Big] \tag{23}$$

Dice loss, on the other hand, focuses on the accuracy of segmented vessel shapes and corrects class imbalance. It is calculated in Equation (24) where smooth is added to prevent zero division.

$$\mathrm{Dice}\big(pr, gt\big) = 1 - \frac{2 \cdot \sum_{i=1}^{N}\big(pr_i \cdot gt_i\big) + \mathrm{smooth}}{\sum_{i=1}^{N} pr_i + \sum_{i=1}^{N} gt_i + \mathrm{smooth}} \tag{24}$$

Finally, the combined loss is formed with a weighting factor $\alpha$ as:

$$\mathrm{CombinedLoss}\big(pr, gt\big) = \alpha \cdot \mathrm{BCE}\big(pr, gt\big) + (1 - \alpha) \cdot \mathrm{Dice}\big(pr, gt\big) \tag{25}$$

In our experiment, a weighting factor of, $\alpha = 0.25$ achieved higher validation metrics. Adam optimizer is used with constant learning rate 0.001 and weight decay of 0.0005. The models are trained for 220 epochs with a dedicated NVIDIA RTX-A4000 GPU.

### 3.2 Evaluation metric

Based on the existing literature review, we have chosen a complete set of evaluation metrics that includes Accuracy($Acc$), Recall($Re$), Specificity($Sp$), Precision($Pr$), $DSC$ and $IoU$. The evaluation metrics, detailed in Equation(26) and (27), include True Positives ($TP$) for correctly identified positive samples, True Negatives ($TN$) for correctly identified negative samples, False Positives ($FP$) for incorrectly identified negative samples, and False Negatives ($FN$) for incorrectly identified positive samples. Moreover, the Area Under the ROC Curve ($AUC$) and Matthews Correlation Coefficient (MCC) are employed to assess the overall performance.

$$\begin{cases} Acc=\dfrac{TP+TN}{TP+FN+FP+TN} \\ Re=\dfrac{TP}{TP+FN} \\ Sp=\dfrac{TN}{TN+FP} \end{cases} \tag{26}$$

$$\begin{cases} Pr=\dfrac{TP}{TP+FP} \\ DSC=\dfrac{2\cdot TP}{2\cdot TP+FP+FN} \\ IoU=\dfrac{TP}{TP+FN+FP} \end{cases} \tag{27}$$

### 3.3 Result evaluation

3.3.1 General comparison　The evaluation of our proposed method was systematically conducted across five distinct datasets. Quantitative metrics derived from this analysis are presented in Tab. 2 through 7, providing a thorough assessment of the model's performance and efficacy. We conducted the evaluation on the two subgroup. Firstly, we trained 3 baseline models-UNet[10], AttUNet[42] and UNet++[13] with the same parameters, loss functions and datasets that are used in our experiment and tested the model following the same evaluation criteria. Referring to the statistical data presented in Tab. 2, it should be noted that the proposed method surpasses all three baseline model across the reported datasets. Although, baseline UNet shows higher recall in CHASE dataset, it achieved lower precision. On the other-hand in IOSTAR dataset, AttnUnet achieved higher precision and specificity with the cost of lower recall. Furthermore, visual evidence in Fig. 6 conclude that the DEFFA-Unet obtains more complete segmentation with lower false-positive alarms. A comparison of the focused red regions in the first and last rows of Fig. 6 reveals that the proposed model is capable of extracting more precise thin vessels than compared methods from low resolution regions with high precision rate.

Additionally, the second row of the Fig. 6 demonstrates the model's ability to effectively delineate vessels within regions of soft exudation. Furthermore, segmentation results in the third row illustrates the model's efficacy in the optic disk region.

Tab. 2　Performance comparison on DRIVE, CHASEDB1, STARE, and IOSTAR: Top performer denoted in red

| | Methods | DRIVE | | | | | | | CHASE | | | | | | |
|---|---|---|---|---|---|---|---|---|---|---|---|---|---|---|---|
| | | $Acc$ | $Re$ | $Sp$ | $AUC$ | $Pr$ | $IoU$ | $DSC$ | $Acc$ | $Re$ | $Sp$ | $AUC$ | $Pr$ | $IoU$ | $DSC$ |
| (A) | UNet[10] | 0.9609 | 0.7583 | 0.9803 | 0.9596 | 0.7879 | 0.6298 | 0.7728 | 0.9623 | 0.8889 | 0.9672 | 0.9764 | 0.6432 | 0.5954 | 0.7464 |
| | AttUNet[42] | 0.9618 | 0.7594 | 0.9813 | 0.9683 | 0.7964 | 0.6360 | 0.7775 | 0.9626 | 0.8834 | 0.9679 | 0.9708 | 0.6470 | 0.5961 | 0.7469 |
| | UNet++[13] | 0.9608 | 0.8100 | 0.9754 | 0.9712 | 0.7600 | 0.6439 | 0.7833 | 0.9725 | 0.8233 | 0.9824 | 0.9838 | 0.7574 | 0.6515 | 0.7889 |
| | Proposed | 0.9693 | 0.8271 | 0.9829 | 0.9861 | 0.8223 | 0.7017 | 0.8247 | 0.9757 | 0.8190 | 0.9862 | 0.9896 | 0.7993 | 0.6793 | 0.8090 |
| | Methods | STARE | | | | | | | IOSTAR | | | | | | |
| | | $Acc$ | $Re$ | $Sp$ | $AUC$ | $Pr$ | $IoU$ | $DSC$ | $Acc$ | $Re$ | $Sp$ | $AUC$ | $Pr$ | $IoU$ | $DSC$ |
| (B) | UNet[10] | 0.9489 | 0.7907 | 0.9612 | 0.9317 | 0.6132 | 0.5276 | 0.6907 | 0.9567 | 0.8329 | 0.9676 | 0.9622 | 0.6938 | 0.6091 | 0.7570 |
| | AttUNet[42] | 0.9605 | 0.7520 | 0.9767 | 0.9410 | 0.7159 | 0.5791 | 0.7335 | 0.9681 | 0.7276 | 0.9892 | 0.9671 | 0.8568 | 0.6487 | 0.7869 |
| | UNet++[13] | 0.9586 | 0.8270 | 0.9688 | 0.9646 | 0.6737 | 0.5905 | 0.7425 | 0.9651 | 0.8158 | 0.9782 | 0.9701 | 0.7678 | 0.6544 | 0.7911 |
| | Proposed | 0.9706 | 0.8728 | 0.9782 | 0.9870 | 0.7576 | 0.6823 | 0.8111 | 0.9690 | 0.8618 | 0.9784 | 0.9849 | 0.7788 | 0.6924 | 0.8182 |

Tab. 3　Performance Comparison with SOTA Methods on DRIVE and STARE: Top performers are denoted in red, green and blue. N. R. denotes not reported metrics by the author

| Methods | DRIVE | | | | | Methods | STARE | | | | |
|---|---|---|---|---|---|---|---|---|---|---|---|
| | Acc | Re | DSC | AUC | MCC | | Acc | Re | DSC | AUC | MCC |
| SFCD[4] | N. R. | 0. 7733 | 0. 7886 | N. R. | 0. 7589 | SFCD[4] | N. R. | 0. 7924 | 0. 7953 | N. R. | 0. 7724 |
| BVS-EL[6] | 0. 9767 | 0. 8173 | N. R. | 0. 9475 | N. R. | BVS-EL[6] | 0. 9791 | 0. 8104 | N. R. | 0. 9813 | N. R. |
| FCNN[8] | 0. 9576 | 0. 8039 | N. R. | 0. 9821 | N. R. | FCNN[8] | 0. 9694 | 0. 8315 | N. R. | 0. 9905 | N. R. |
| DUNET[3] | 0. 9566 | 0. 7963 | 0. 8203 | 0. 9802 | N. R. | DUNET[3] | 0. 9641 | 0. 7595 | 0. 8143 | 0. 9832 | N. R. |
| NUPL[14] | 0. 9512 | 0. 8060 | 0. 7863 | 0. 9748 | N. R. | NUPL[14] | 0. 9641 | 0. 8230 | 0. 7947 | 0. 9620 | N. R. |
| CSGNet[19] | 0. 9576 | 0. 7943 | 0. 8310 | 0. 9823 | N. R. | CSGNet[19] | 0. 9692 | 0. 8298 | **0. 8493** | 0. 9895 | N. R. |
| HANET[24] | 0. 9581 | 0. 7991 | 0. 8293 | 0. 9823 | N. R. | HANET[24] | 0. 9673 | 0. 8186 | 0. 8379 | 0. 9881 | N. R. |
| CSUNet[22] | 0. 9565 | 0. 8071 | 0. 8251 | 0. 9801 | N. R. | CSUNet[22] | 0. 9702 | **0. 8432** | 0. 8516 | 0. 9825 | N. R. |
| DPFNet[23] | 0. 9570 | 0. 8279 | **0. 8303** | **0. 9824** | N. R. | DPFNet[23] | 0. 9655 | 0. 8287 | 0. 8366 | **0. 9898** | N. R. |
| ATSM[26] | 0. 9538 | 0. 7631 | N. R. | 0. 9750 | N. R. | ATSM[26] | 0. 9638 | 0. 7735 | N. R. | 0. 9833 | N. R. |
| MPCEM[25] | 0. 9574 | 0. 8083 | N. R. | 0. 9822 | **0. 7984** | MPCEM[25] | 0. 9695 | 0. 8162 | N. R. | **0. 9898** | 0. 8066 |
| Proposed | **0. 9693** | **0. 8271** | 0. 8247 | 0. 9861 | 0. 8079 | Proposed | **0. 9706** | 0. 8728 | 0. 8111 | 0. 9870 | **0. 7977** |

Tab. 4　Performance comparison with SOTA methods on CHASEDB1 and IOSTAR: Top performers are denoted in red, green and blue. N. R. denotes not reported metrics by the author

| Methods | CHASE | | | | | Methods | IOSTAR | | | | |
|---|---|---|---|---|---|---|---|---|---|---|---|
| | Acc | Re | DSC | AUC | MCC | | Acc | Re | DSC | AUC | MCC |
| SFCD[4] | N. R. | 0. 7303 | 0. 7202 | N. R. | 0. 6928 | SFCD[4] | N. R. | 0. 7924 | 0. 7953 | N. R. | 0. 7724 |
| FCNN[8] | 0. 9653 | 0. 7779 | N. R. | 0. 9855 | N. R. | GUNet[9] | 0. 9641 | N. R. | 0. 8109 | 0. 9872 | N. R. |
| DUNet[3] | 0. 9724 | 0. 8229 | 0. 7853 | 0. 9863 | N. R. | DRNet[12] | 0. 9713 | 0. 8082 | N. R. | **0. 9873** | **0. 8017** |
| CSGNet[19] | 0. 9681 | 0. 7947 | **0. 8247** | **0. 9886** | N. R. | CSGNet[19] | 0. 9659 | 0. 8076 | **0. 8288** | 0. 9875 | N. R. |
| HANET[24] | 0. 9670 | 0. 8239 | 0. 8191 | 0. 9871 | N. R. | HANET[24] | 0. 9652 | 0. 7538 | 0. 8161 | 0. 9859 | N. R. |
| SCS-Net[45] | **0. 9744** | **0. 8365** | N. R. | 0. 9867 | N. R. | SCS-Net[45] | **0. 9706** | **0. 8255** | 0. 8516 | 0. 9865 | N. R. |
| DPFNet[23] | 0. 9676 | **0. 8303** | 0. 8302 | 0. 9868 | N. R. | NFN+[18] | 0. 9683 | 0. 7921 | N. R. | 0. 9803 | N. R. |
| MPCEM[25] | 0. 9654 | 0. 8138 | N. R. | 0. 9850 | 0. 7969 | CSNet[17] | 0. 9627 | N. R. | 0. 7984 | 0. 9855 | N. R. |
| Proposed | 0. 9757 | 0. 8190 | 0. 8090 | 0. 9896 | **0. 7962** | Proposed | 0. 9690 | 0. 8618 | 0. 8182 | 0. 9849 | 0. 8025 |

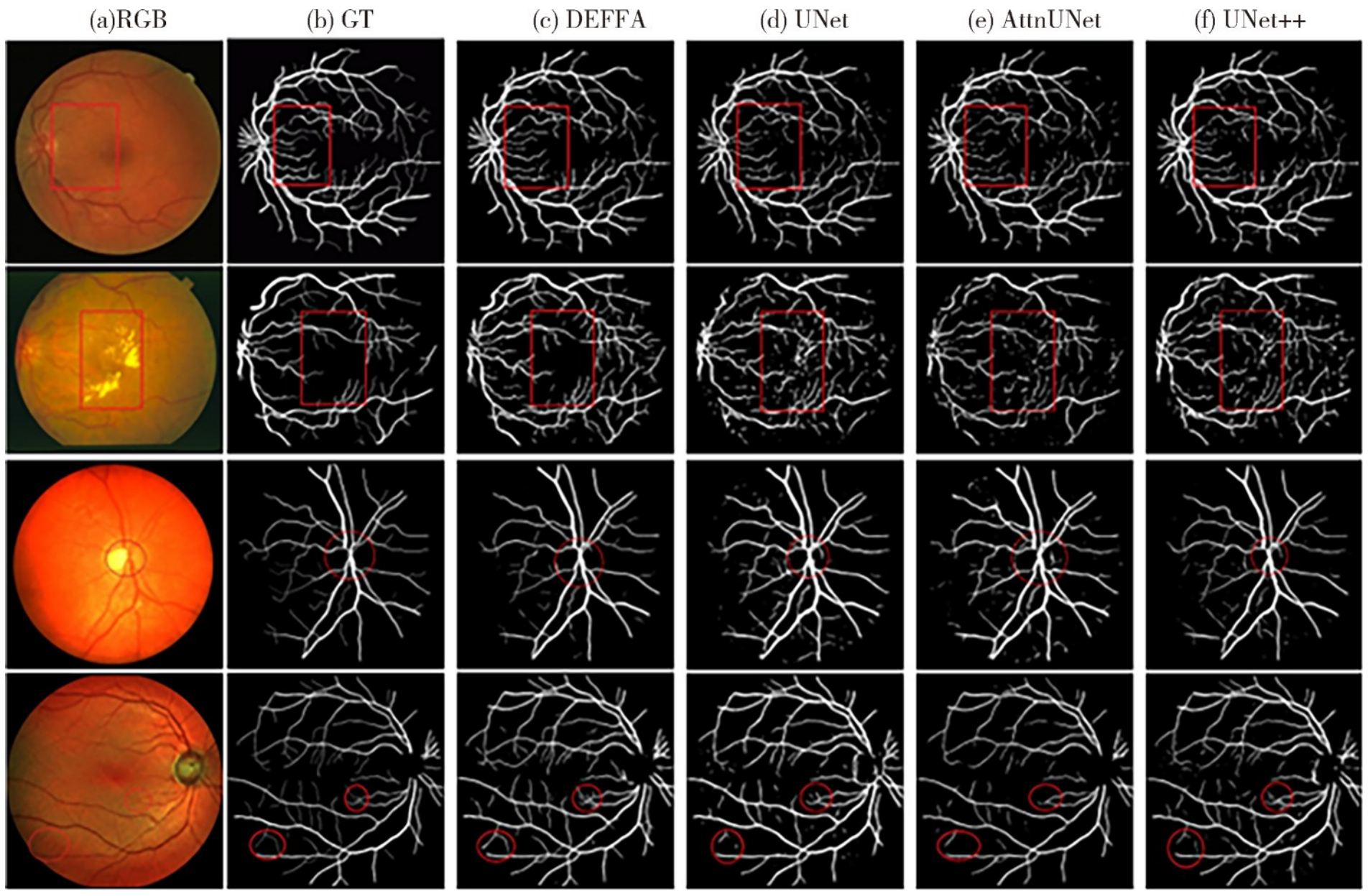

Fig. 6　Visual results comparison on four images from DRIVE, STARE, CHASEDB1 and IOSTAR(from top to the bottom row)

Secondly, we compare our methods with current state-of-the-art (SOTA) methods as shown in Tab. 3 and Tab. 4, across six distinct categories: fully convolutional neural network based method-FCNN[8], feature-guided ensemble learning methods-SFCD[4] and BVS-EL[6], advanced UNet-based methods DUNET[3] and NUPL[14], cascaded UNet based method- CSGNet[19], dual-encoding based methods- CSUNet[22], DPFNet[23], NFN+[18], and multi task and multi-modal based methods-ATSM[26] and MPCEM[25]. Overall, our proposed method achieved the best scores in all reported metrics as shown in Tab. 3 and Tab. 4. Especially, the proposed DEFFA-UNet establishes a new standard on the DRIVE dataset, achieving the highest scores in both AUC and MCC, along with the second highest scores in accuracy and recall. This performance underscores its robustness in retinal vessel segmentation, a superiority that is further corroborated by its performance on the CHASEDB1 dataset, where it attained the highest scores in AUC and accuracy, and the second highest in MCC. Furthermore, statistical evidence from Tab. 3 and Tab. 4 on the STARE and IOSTAR datasets demonstrates that the proposed method not only achieved the top position in recall but also achieved best scores in MCC, accuracy and DSC with competitive AUC scores which highlights its superior capability in robust vessel segmentation. Notably, the model's ability to outperform dual-encoding methods such as CSUNet[22] and DPFNet[23] in all compared datasets reinforces its efficacy over these established similar approaches.

However, although the CSUNet[22] achieved a marginally higher DSC on the STARE dataset, our model maintained a lead in AUC, alongside superior accuracy and MCC, indicative of its overall reliability. In a comparative context, DEFFA-Unet surpasses traditional fully convolutional network (FCNN) methods[8], ensemble-based approaches[4,6], as well as advanced UNet variants[3,14], and multi-task frameworks [25, 26]. Such a comprehensive outperformance across all evaluation criteria point to the DEFFA-Unet's comprehensive capability to enhance retinal vessel segmentation tasks across varied datasets and conditions.

Tab. 5　Cross-domain generalization with SOTA Methods on STARE, CHASEDB1 and HRF: Top performers are denoted in red, green and blue. N. R. denotes not reported metrics by the author

| Methods | Trained on | Tested on | | | | | | | | |
|---|---|---|---|---|---|---|---|---|---|---|
| | | STARE(S) | | | CHASE(C) | | | HRF(H) | | |
| | | DSC | AUC | MCC | DSC | AUC | MCC | DSC | AUC | MCC |
| UNet[10] | DRIVE(D) | 0.7379 | 0.9607 | 0.7214 | 0.7080 | 0.9520 | 0.68774 | 0.6416 | 0.9395 | 0.6295 |
| UNet++[13] | DRIVE(D) | 0.7447 | 0.9726 | 0.7319 | 0.7168 | 0.9643 | 0.7003 | 0.6418 | 0.9542 | 0.6362 |
| AttUNet[42] | DRIVE(D) | 0.7459 | 0.9724 | 0.7290 | 0.7126 | 0.9665 | 0.6926 | 0.6569 | 0.9491 | 0.6444 |
| W-Net[20] | DRIVE(D) | **0.7976** | **0.9828** | **0.7765** | **0.7649** | **0.9756** | **0.7402** | **0.7112** | 0.9612 | 0.6886 |
| HGC-Net[46] | DRIVE(D) | N. R. | N. R. | N. R. | 0.7490 | N. R. | 0.7290 | N. R. | N. R. | N. R. |
| DPF-Net[23] | DRIVE(D) | N. R. | 0.9735 | N. R. | N. R. | N. R. | N. R. | N. R. | N. R. | N. R. |
| DoFE[37] | Leave-1(D,S,C,H) | N. R. | 0.9725 | N. R. | N. R. | 0.9628 | N. R. | N. R. | 0.8948 | N. R. |
| AADG [36] | Leave-1(D,S,C,H) | 0.8179 | 0.9796 | N. R. | 0.7834 | 0.9695 | N. R. | 0.7257 | 0.9193 | N. R. |
| Baseline | DRIVE(D) | 0.7774 | 0.9718 | 0.7655 | 0.7432 | 0.9497 | 0.7258 | 0.7039 | **0.9744** | 0.7067 |
| Proposed | DRIVE(D) | 0.7918 | 0.9889 | 0.7814 | 0.7526 | 0.9784 | 0.7365 | 0.7135 | 0.9786 | **0.7056** |

3.3.2　Model Generalization　In cross-domain evaluation, the DEFFA-Unet model demonstrated exceptional generalization across the STARE, CHASEDB1, and HRF datasets with the significant higher scores in all considered metrics as shown in Tab. 5. Compared with W-Net[20] and HGC-Net[45] that used both the source and target dataset for model generalization, our method achieved higher AUC and MCC in all three datasets without any fine-tuning with target datasets. Notably, proposed model's performance on the HRF dataset is particularly commendable. Despite the inherently challenging nature of this dataset, the model secures competitive evaluation scores in DSC and the second-highest MCC with the best scores in AUC further solidifying its performances. Despite not being trained on multi-domain data, the DEFFA Unet model was bench-marked against prominent multi-domain approaches such as DoFE [37] and AADG[36], which utilized a leave-one-out training methodology across four datasets. Remarkably, the DEFFA-Unet model consistently exhibited superior results in this comparison, highlighting its exceptional adaptability and robustness in handling data from various sources without compromising on the segmentation quality.

Tab. 6　Cross-domain generalization with Baseline Methods on STARE, CHASEDB1 and DRIVE: Top performers are denoted in red. P denotes the parameter and w/o D-I denotes without domain-invariant feature guidance.

| Trained on IOSTAR Methods | P(MB) | Tested on STARE DSC | STARE AUC | STARE MCC | CHASE DSC | CHASE AUC | CHASE MCC | DRIVE DSC | DRIVE AUC | DRIVE MCC |
|---|---|---|---|---|---|---|---|---|---|---|
| UNet[10] | 7.42 | 0.5986 | 0.8912 | 0.5767 | 0.7140 | 0.9537 | 0.6944 | 0.6625 | 0.9096 | 0.6377 |
| UNet++[13] | 8.75 | 0.7149 | 0.9287 | 0.6968 | 0.7522 | 0.9629 | 0.7348 | 0.7271 | 0.9383 | 0.7069 |
| AttUNet[42] | 7.51 | 0.7565 | 0.9709 | 0.7402 | 0.7543 | 0.9762 | 0.7371 | 0.7616 | 0.9710 | 0.7427 |
| Proposed w/o CSA | 2.85 | 0.7578 | 0.9415 | 0.7444 | 0.7690 | 0.9749 | 0.7525 | 0.7743 | 0.9594 | 0.7594 |
| Proposed w/o D-I | 2.85 | 0.7572 | 0.9768 | 0.7405 | 0.7662 | 0.9819 | 0.7530 | 0.7706 | 0.9720 | 0.7511 |
| Proposed | 2.85 | 0.7847 | 0.9834 | 0.7699 | 0.7726 | 0.9831 | 0.7588 | 0.7738 | 0.9771 | 0.7548 |

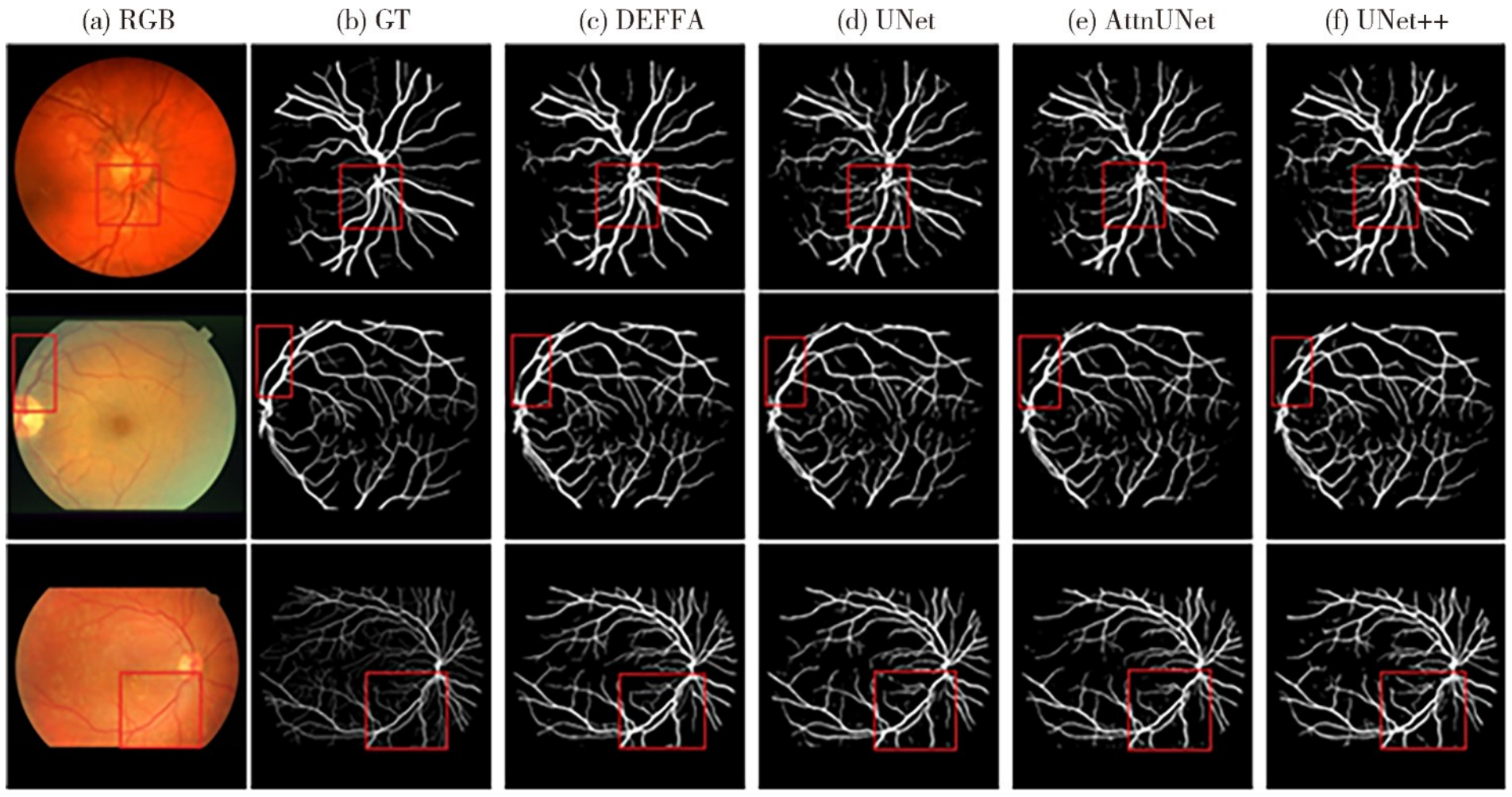

Fig. 7　Visual cross-validation results comparison on three images from CHASEDB1, STARE and HRF(from top to the bottom row)

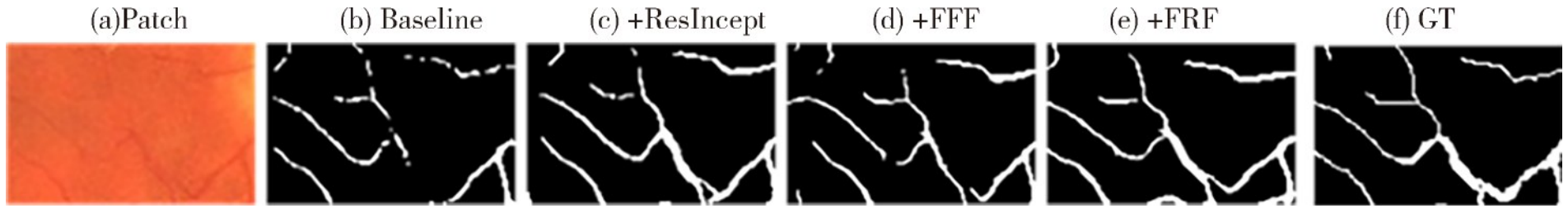

Fig. 8　Visual Ablation study results on a DRIVE Image patch

We also conducted the cross-modality test on a recently developed Scanning Laser Ophthalmoscopy (SLO) dataset IOSTAR[33]. We trained the model with IOSTAR and tested on STARE, CHASEDB1 and DRIVE datasets. The cross validation results in Tab. 6 reveal that proposed method consistently outperforms the benchmark models in all three target datasets with the smaller parameter footprint. It should be noted that, in Tab. 6, 'Prop. w/o D-I' refers to the model trained without domain-invariant feature guidance which highlights the significance of the proposed hybrid feature encoding method. In addition, comparison without SOTA-CSA outline that CSA also collectively en-hance the model generalization. Fig. 7 showcases the cross-validation outcomes of DEFFA-Unet against three baseline models, illustrating a comprehensive segmentation capability. Notably, a distinctive strength of proposed method is its ability to maintain the connectivity of central vessels, a common challenge for many existing models. As highlighted by the red annotation regions, where most approaches struggle to preserve these vital vessel connections, our DEFFA-Unet demonstrates a consistent capability.

3.3.3 Ablation Study　In the ablation study, the benefits derived from each component within the proposed model are shown in Tab. 7. In the table the baseline is dual-encoder UNet without res-inception (ResIncept) decoder block, feature filtering fusion (FFF) module and feature reconstructing fusion (FRF) module. The sequential integration of ResIncept, FFF module and FRF module demonstrates consistent improvements in the DSC, AUC, and MCC metrics, which signifies the effectiveness of each module. Notably, the FRF module ensured higher precision, aligning with our research priority of minimizing false-positive cases. In addition, visual analysis presented in Fig. 8 demonstrates that the proposed model consistently enhances vessel connectivity and generates detailed segmentation maps, attributable to the systematic integration of each module, underscoring the model's architectural efficacy in improving retinal blood vessel segmentation. Furthermore, the ablation study for data balancing and data augmentation precisely point out the effectiveness of proposed SOTA-CSA and JESB methods.

Tab. 7　Ablation study on the DRIVE Dataset. Baseline is the Dual-encoder little UNet

| Modules | Pr. | DSC | AUC | MCC |
|---|---|---|---|---|
| Baseline | 0.8177 | 0.7985 | 0.9445 | 0.7803 |
| +ResIncept decoder | 0.7953 | 0.8108 | 0.9786 | 0.7930 |
| +FFF | 0.8057 | 0.8200 | 0.9857 | 0.8026 |
| +FRF(DEFFA-Unet) | 0.8223 | 0.8247 | 0.9861 | 0.8079 |
| Without SOTA-CSA | 0.7872 | 0.8219 | 0.9860 | 0.8050 |
| Without JESB | 0.7687 | 0.8179 | 0.9858 | 0.8013 |

## 4 Discussion

Significant advancements have been achieved in the field of deep learning-based computer-aided detection systems (CADs) over the years, particularly in the area of retinal vessel segmentation. Challenges like fine-vessel segmentation[21,23] and generalized model development[20,36], with the limited annotated class imbalance data remains. Considering the fact, proposed method focuses on improving fine vessel segmentation with boosted generalized capability by integrating new methods for data balancing and data augmentation with a comprehensive hybrid feature encoding method. Our method can effectively handle fine-vessel segmentation with enhanced generalization capabilities. To enrich feature encoding, our proposed method employs a feature-guided dual encoding approach. Compared to existing similar methods, such as DPFNet[23] and CSUNet[22], our approach demonstrates superior segmentation and generalization capabilities. We attribute the success of our method to meticulous data processing and careful model design. By integrating proposed data balancing and augmentation techniques, our model, trained with a hybrid feature encoding method, processes both domain-specific and domain-invariant features, unlike the aforementioned methods which focus solely on domain-specific features, potentially restricting their generalization capabilities. Moreover, while both DPFNet and CSUNet utilize a progressive fusion module that ensures refined feature propagation at the expense of transmitting irrelevant low-level features to deeper layers, our model employs a Feature Filtering Fusion FFF module before the bottleneck to ensure robust feature passing to the deepest layers. It is crucial to clarify that this paper does not claim superiority of the proposed method over others. Instead, our aim is to underscore the effectiveness of the dual encoding approach coupled with domain-based feature encoding techniques, which enhance both extensive feature extraction and model generalization. This nuanced approach addresses the critical challenge of developing both robust and generalized segmentation models effectively. However, still there are some failure case to be considered. the model's ability in vessel connectivity retaining from some low-contrast images is not up to the mark. Additionally, the evaluation metrics derived from STARE and IOSTAR datasets in model evaluation highlight the noteworthy imbalance in recall and precision metrics. This disparity suggests the proposed model may prioritize sensitivity over precision in these two datasets. Although, the $F1$ score in these two dataset evaluation is in competitive range, our future efforts would focus on enhancing model calibration techniques or incorporating balanced loss functions to mitigate this issue, striving for a more harmonized precision-recall trade-off. Additionally, integrating an automated contrast adjustment methods in high-frequency feature extraction method can be explored in the future work to address low contrast issue.

Furthermore, to enhance vessel segmentation in challenging clinical set-up, priority should be given on expanding diverse image datasets and improving the image acquisition quality following methods like [47]. In addition, a standard evaluation protocols should be followed in accessing the model performances that avoid misleading metrics like accuracy in favor of more informative measures. We believe such protocols with improved image acquisition quality will lead to more reliable computer aided segmentation system.

## 5 Conclusions

In our research, we present an innovative feature- guided dual encoding UNet architecture designed for the segmentation of retinal blood vessels, which notably advances segmentation accuracy and model generalization. Our method leverages domain-invariant features alongside domain-specific ones to markedly enhance performance on novel datasets. At the core of our methodology, we integrate two innovative fusion modules and a modified res-inception decoder into the proposed hybrid feature encoding dual-encoder Unet. By incorporating domain-guided richer feature extraction, the proposed Feature Filtering Fusion (FFF) module, Feature Reconstructing Fusion (FRF) module, and res-inception decoder can significantly enhance vessel connectivity with higher precision, mitigate semantic gap in feature reconstruction, and greatly improve the model's generalization capability on unseen data. Additionally, we introduce a novel data augmentation strategy utilizing SOTA color space transformations, significantly boosting segmentation effectiveness and model generalization. Collectively the proposed met-hod achieves superior segmentation performance with enhanced model generalization than current mainstream methods. This work underscores the importance of feature-guided approaches and advanced feature fusion in improving model performance and generalization in medical imaging. Future work will focus on improving feature learning in areas with lesions and creating an automatic pipeline for processing domain-invariant features.